\documentclass[fleqn,usenatbib]{mnras}

\usepackage{newtxtext,newtxmath}

\usepackage[T1]{fontenc}
\usepackage{ae,aecompl}

\usepackage{graphicx}	
\usepackage{amsmath}	
\usepackage{amssymb}	

\usepackage{hyperref}
\usepackage{multirow}
\usepackage{xcolor}

\newcommand{\HI}{H{\sc\ i}}
\newcommand{\HII}{H{\sc\ ii}}
\newcommand{\OI}{O{\sc\ i}}
\newcommand{\SiII}{Si{\sc\ ii}}
\newcommand{\SiIII}{Si{\sc\ iii}}
\newcommand{\SiIV}{Si{\sc\ iv}}
\newcommand{\CIV}{C{\sc\ iv}}

\newcommand{\CII}{C{\sc\ ii}}
\newcommand{\SII}{S{\sc\ ii}}
\newcommand{\FeII}{Fe{\sc\ ii}}
\newcommand{\PII}{P{\sc ii}}

\newcommand{\msun}{M_\odot}
\newcommand{\mmsun}{$\msun$}
\newcommand{\kms}{{\rm km \, s^{-1}}}
\newcommand{\mkms}{km s$^{-1}$}

\newcommand{\vlsr}{v_{\rm LSR}}
\newcommand{\mvlsr}{$\vlsr$}

\defcitealias{fox13}{Fox13}
\defcitealias{fox14}{Fox14}

\title[CGM of WLM]{Tentative Detection of the Circumgalactic Medium of the Isolated Low-Mass Dwarf Galaxy WLM}

\author[Y. Zheng et al.]{Yong Zheng$^{1, 2}$\thanks{E-mail: yongzheng@berkeley.edu}, 
Mary E. Putman$^{3}$, 
Andrew Emerick$^{3, 4}$,
Kristen B. W. McQuinn$^{5, 6}$, 
\newauthor
Jessica K. Werk$^{7}$, 
Felix J. Lockman$^{8}$, 
Benjamin D. Oppenheimer$^{9}$, 
\newauthor
Andrew J. Fox$^{10}$, 
Evan N. Kirby$^{11}$,
Joseph N. Burchett$^{12}$
\\
$^{1}$ Department of Astronomy, University of California, Berkeley, Berkeley, CA 94720, USA\\
$^{2}$ Miller Institute for Basic Research in Science, University of California, Berkeley, Berkeley, CA 94720, USA\\
$^{3}$ Department of Astronomy, Columbia University, New York, NY 10027, USA\\
$^{4}$ Department of Astrophysics, American Museum of Natural History, New York, NY, USA \\
$^{5}$ The University of Texas at Austin, Department of Astronomy, Austin, Texas 78712, USA, \\
$^{6}$ Rutgers University, Department of Physics and Astronomy, New Jersey, NJ 08854, USA\\
$^{7}$ Department of Astronomy, University of Washington, Seattle, WA 98195-1580, USA\\
$^{8}$ National Radio Astronomy Observatory, Green Bank, WV 24944, USA\\
$^{9}$ Center for Astrophysics and Space Astronomy, Department of Astrophysical and Planetary Sciences,
\\
  University of Colorado, Boulder, CO 80309, USA\\
$^{10}$ AURA for ESA, Space Telescope Science Institute, Baltimore, MD, USA\\
$^{11}$ Astronomy Department, California Institute of Technology, Pasadena, CA 91125, USA \\
$^{12}$ Department of Astronomy, University of California, Santa Cruz, Santa Cruz, CA, USA 95064
}

\date{Accepted 2019 September 6. Received 2019 July 15; in original form 2019 April 3}

\pubyear{2019}

\begin{document}
\label{firstpage}
\pagerange{\pageref{firstpage}--\pageref{lastpage}}
\maketitle

\begin{abstract}
We report a tentative detection of the circumgalactic medium (CGM) of WLM, an isolated, low-mass (log$M_*/M_\odot\approx7.6$), dwarf irregular galaxy in the Local Group (LG). We analyze an HST/COS archival spectrum of a quasar sightline (PHL2525), which is 45 kpc (0.5 virial radius) from WLM and close to the Magellanic Stream (MS). Along this sightline, two ion absorbers are detected in Si~II, Si~III, Si~IV, C~II, and C~IV at velocities of $\sim$-220 km s$^{-1}$ (Component v-220) and $\sim$-150 km s$^{-1}$ (Component v-150). To identify their origins, we study the position-velocity alignment of the components with WLM and the nearby MS. Near the Magellanic longitude of PHL2525, the MS-related neutral and ionized gas moves at $\lesssim-190$ km s$^{-1}$, suggesting an MS origin for Component v-220, but not for Component v-150. Because PHL2525 passes near WLM and Component v-150 is close to WLM's systemic velocity ($\sim$-132 km s$^{-1}$), it is likely that Component v-150 arises from the galaxy's CGM.  This results in a total Si mass in WLM's CGM of $M_{\rm Si}^{\rm CGM}\sim(0.2-1.0)\times10^5~M_\odot$ using assumption from other COS dwarf studies. Comparing $M_{\rm Si}^{\rm CGM}$ to the total Si mass synthesized in WLM over its lifetime ($\sim$1.3$\times10^5~M_\odot$), we find $\sim$3\% is locked in stars, $\sim$6\% in the ISM, $\sim$15\%-77\% in the CGM, and the rest ($\sim$14\%-76\%) is likely lost beyond the virial radius. Our finding resonates with other COS dwarf galaxy studies and theoretical predictions that low-mass galaxies can easily lose metals into their CGM due to stellar feedback and shallow gravitational potential.

\end{abstract} 




\section{Introduction} 
\label{sec:intro}

At redshift $\sim0$, only $\sim10-20$\% of the baryons predicted by $\Lambda$ cold dark matter ($\Lambda{\rm CDM}$) cosmology \citep{planck16} are found in stars and the interstellar medium (ISM) of galaxies \citep{persic92, mcgaugh10, behroozi10, peeples14}. In addition, hot gas in clusters and groups, as detected in X-ray, contributes $\sim4\%$ of the predicted baryons \citep{fukugita04, bregman07}. The rest of the baryons, a.k.a the missing baryons, are likely to reside in the circumgalactic medium (CGM) of galaxies, and the surrounding intergalactic medium \citep{bregman07, werk14, danforth16, tumlinson17, shull17}. So far, baryons in the CGM have been detected using quasar (quasi-stellar object; QSO) absorption line diagnostics for a wide range of galaxy masses (e.g., dwarfs, $L\sim L_*$, and luminous red galaxies; for a non-exhaustive list: \citealt{tumlinson11, rudie12, werk14, bordoloi14, liang14, burchett16, johnson17, smailagic18}). Emission line mapping of Lyman-$\alpha$ photons near star-forming galaxies and bright QSOs have also found large halos spanning a few tens to hundreds kpc, which could be massive baryonic reservoirs \citep{cantalupo14, hennawi15, borisova16, cai17}.

Among all the galaxies being probed, low-mass dwarf galaxies are predicted to be the least likely to retain their metals in stars or ISM due to their shallow gravitational potential. \cite{maclow99} show that low-mass dwarf galaxies could lose the majority of their synthesized metals to the CGM and intergalactic medium (IGM) because of supernova (SN) feedback. 
\cite{christensen18} also suggest that for galaxies with stellar mass $M_*\sim10^{7}~\msun$, 85\% of their synthesized metals do not remain in the galaxies by $z = 0$. 
These metals are mostly transported into the CGM and beyond the virial radius through outflows. In particular, for galaxies with $M_*\lesssim 10^7~\msun$, the outflowing metals are mostly trapped in the CGM probably due to the weakening of star-formation activities (see figure 3 in \citealt{christensen18}). \cite{muratov17} show that for galaxies at $M_*\sim10^{7-7.6}~\msun$, 50\% of all the metals (by mass) reside in the CGM, more than 90\% of which is in cool phase with $T\sim10^{4.0-4.7}$ K (see also \citealt{ma16}). Results from different simulations vary quantitatively, probably due to different treatment of feedback recipes (see \citealt{hu17, emerick18, emerick19} for a detailed work on the impact of high-resolution stellar feedback modeling in regulating the efficiency of outflows). Overall, they agree that, for dwarf galaxies with 6$<$log$(M_*/M_\odot)<10$, a large fraction of metals are lost into their CGM, if not further into the IGM. 


Observationally, a metal deficit has been reported in a number of low-mass dwarf galaxies in the local universe. For example, \cite{kirby11} study the metal mass contained in stars of eight gas-poor dwarf spheroidal (dSph) galaxies over stellar mass ranges of $10^{5.7-7.3}~\msun$, and find that the galaxies have commonly lost 96\% to more than $99\%$ of the metals synthesized. \cite{mcquinn15} find that Leo P,  an isolated gas-rich dwarf galaxy with $M_*\sim10^{5.7}~\msun$, has lost 95\% of the oxygen produced throughout its star-formation history. In addition, the QSO absorption line technique has been employed by a number of authors in search of lost metals in the CGM of dwarf galaxies. \cite{bordoloi14} conduct an HST/COS search of metals in the CGM of 43 low-mass galaxies ($M_*\approx 10^{8.2-10.2}~\msun$) at $z\leq0.1$. They find a covering fraction of $\sim40\%$ for \CIV\ within the virial radius ($R_{\rm vir}$), and a minimum carbon mass of $1.2\times10^6~\msun$ within 0.5$R_{\rm vir}$ (see also \citealt{liang14, burchett16}). 

In this article, we present our effort in searching for the CGM of Wolf-Lundmark-Melotte (WLM), a nearby, gas-rich, low-mass dwarf irregular (dIrr) galaxy. WLM moves at a heliocentric velocity of $-130$ \mkms\ \citep{jackson04} and $\vlsr=-132~\kms$ with respect to the Local Standard of Rest (LSR). It is located on the outskirts of the Local Group at a distance of $0.93\pm0.03$ Mpc \citep{mcconnachie05, mcconnachie12} from the Milky Way. WLM is an isolated galaxy $\sim210$ kpc away from its nearest neighbor (the Cetus dwarf spheroidal galaxy; \citealt{whiting99}). The virial radii ($R_{\rm vir}$) of WLM and Cetus are $\sim90$ kpc and $\sim60$ kpc, respectively (see below for $R_{\rm vir}$ definition); therefore, their halos do not overlap. In addition, \cite{leaman12} find that the stellar kinematics of WLM show no sign of tidal influence, further supporting its isolation within the Local Group. Therefore, WLM is an excellent candidate to study metal distributions due to stellar feedback without apparent influence from the environment. 




WLM has a stellar mass of $M_*=4.3\times10^7~\msun$ \citep{jackson07, mcconnachie12}, and an \HI~ mass of $M_{\rm HI}=(6.3\pm0.3)\times10^7~\msun$ (\citealt{kepley07}; see also \citealt{huchtmeier81, barnes04, jackson04}). \cite{leaman12} estimate WLM's dark-matter halo mass $M_{\rm h}$ based on its stellar rotation curve and line-of-sight dispersion velocity. They find $M_{\rm h}({\rm ISO})=(2.6\pm0.2)\times10^{10}~\msun$ assuming an isothermal spherical halo, and $M_{\rm h}({\rm NFW})=(8.9\pm0.8)\times10^9~\msun$ if using an NFW profile instead. The virial radius $R_{\rm vir}$, defined as the radius within which the mean density is 200 times the cosmic critical density $\rho_{\rm c}$, is $R_{\rm 200}({\rm ISO})=60.5$ kpc and $R_{\rm 200}({\rm NFW})=42.0$ kpc, respectively (see their table 3). Meanwhile, a different definition of virial radius is also often adopted, which is referred to 200 times the matter density $\rho_{\rm m}\equiv \rho_{\rm c}\Omega_{\rm m}$ (e.g., \citealt{werk13, bordoloi14, shull14}). In such cases, $R_{200}$ is systemically $(\Omega_{\rm m})^{-1/3}\approx1.48$ times higher than the one defined with $\Omega_{\rm c}$\footnote{We adopt $\Omega_{\rm m}=0.308$ from \cite{planck16} cosmological parameters.}. We note that, in any case, the derivation of $R_{\rm 200}$ is highly uncertain due to uncertainties and scatter in the $M_*-M_h$ relation (e.g., \citealt{moster10, behroozi10}). For instance, the COS-Halos survey \citep{tumlinson13} quote an uncertainty of 50\% in $R_{\rm 200}$ due to errors propagated from the $M_*-M_h$ relation. In this work, we adopt the virial radius of WLM as $R_{200}^{\rm WLM}\equiv R_{\rm 200}({\rm ISO})*\Omega_{\rm m}^{-1/3}=89.7$ kpc, which is defined with respect to the critical matter density in an isothermal spherical halo. This definition is chosen to be consistent with the one used by the COS-Dwarfs study (equation 1; \citealt{bordoloi14}), such that we can make a direct comparison between our analysis and their CGM observations (see \S\ \ref{sec:discuss}). 



WLM has an iron abundance of [Fe/H]$=-1.28\pm0.02$ dex from spectroscopic studies of red giant branch stars \citep{leaman13}, and a gas-phase oxygen abundance of 12+log(O/H)$_{\rm WLM}=7.83\pm0.06$ \citep{lee05}. Its current day star-formation rate is $2.7\times10^{-4}~\msun~ {\rm yr}^{-1}$ as averaged over the galaxy \citep{dolphin00}. Star formation history (SFH) analysis on selected HST/WFPC2 fields in WLM shows that the galaxy has experienced an active star formation epoch $1-3$ Gyrs ago \citep{dolphin00, weisz14a}. 
Escaping metals, carried by outflows driven by stellar feedback, may pollute the galaxy's CGM over Gyr timescales, leaving traces of elements potentially detectable in ultraviolet (UV) absorption lines.

\begin{figure*}
\includegraphics[width=2\columnwidth]{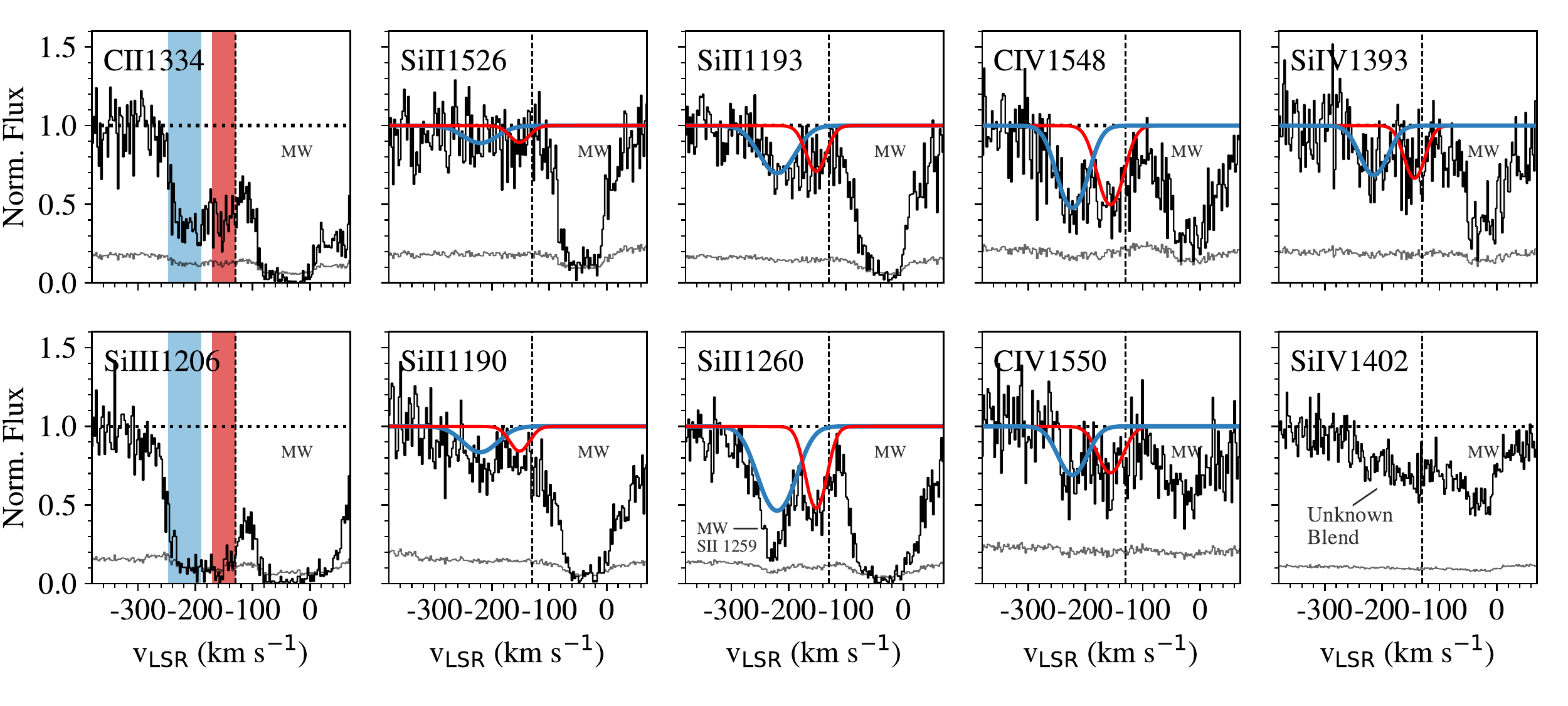}
\caption{Detection of ion absorption lines toward PHL2525 and the corresponding Voigt-profile fits. For ions with multiple transitions, we conduct two-component fits simultaneously to all the available lines. We do not attempt to decompose the \CII~1334 \AA\ and \SiIII~1206 \AA\ lines as each of them only have one transition and appears saturated. We instead measure the column densities for regions highlighted in blue and red. For each transition, we refer to the component at $\vlsr\sim-150~\kms$ (red) as Component v-150, and the one at $\vlsr\sim-220~\kms$ (blue) as Component v-220. We note that \SiII\ 1260 is blended with \SII\ 1259 from the MW's ISM, which has been taken into account when we run the Voigt-profile fitting. See \S \ref{sec:cos} for more details.}
\label{fig:cosspec}
\end{figure*}

This article focuses on UV and \HI~21cm observation along a QSO sightline, PHL2525, through the CGM of WLM. In \S\ \ref{sec:data}, we present the archival UV spectrum of PHL2525 retrieved from the Mikulski Archive for Space Telescopes (MAST), and follow-up \HI~ data obtained with the Green Bank Telescope (GBT). In \S\ \ref{sec:origin}, we investigate the origins of the UV detection based on position-velocity alignment. In \S\ \ref{sec:discuss}, we discuss the implication on WLM's CGM should the detected gas originate from the galaxy. We summarize in \S\ \ref{sec:sum}. The codes and data used in this work can be found on \href{https://github.com/yzhenggit/Zheng19_WLMCGM.git}{GitHub: yzhenggit/Zheng19\_WLMCGM.git}. 

\section{Data Reduction and Measurements}
\label{sec:data}

\subsection{HST/COS Spectral Analysis}
\label{sec:cos}

We retrieve an HST/COS spectrum of QSO PHL2525 from MAST. The sightline is at ${\rm R.A.=00h00m24.42s}$ (0.1018 degree), ${\rm DEC=-12d45m47.76s}$ (-12.7632 degree), and at an impact parameter of $45.2$ kpc (0.5$R_{200}^{\rm WLM}$) from WLM. The spectrum was obtained as part of GO Program \href{http://www.stsci.edu/cgi-bin/get-proposal-info?id=12604&observatory=HST}{12604} (PI: A. Fox) with 2146/2772 seconds of exposure in G130M/G160M gratings, reaching a signal-to-noise ratio of 22/19, respectively (see table 1 in \citealt{fox13}; \citetalias{fox13} hereafter). The spectrum has a resolution of $R\approx16,000$, corresponding to an instrumental velocity resolution of ${\rm FWHM}\approx19$ \mkms. To coadd the multiple x1d.fits files provided by MAST, we adopt a spectral coadding IDL code {\it coadd\_x1d.pro} developed by \cite{danforth10}. The code works as follows:  First, it randomly selects one exposure as the reference and cross-correlates the remaining exposures using strong ISM absorption lines over spectral regions of 10~\AA. Once any wavelength offset is resolved among the exposures, the offset is applied to all the spectra to line up with the reference exposure wavelength. Finally, the code generates a coadded spectrum that contains exposure-weighted average flux and exposure-weighted, inverse-invariance averaged error array. We also visually inspect line features from each exposure before coaddition to make sure that the corresponding features are not due to instrumental artifacts intrinsic to the spectrograph (\citealt{coshandbook}; COS Data Handbook, V4.0). In the following, we analyze and display the spectra in their native spectral resolution; we do not re-bin the data. 


\begin{table}
\centering
\caption{\textbf{Ion Line Measurements}. \textbf{Col (1)}: Column density. For \CIV, \SiII, and \SiIV, log$N$ is measured from Voigt-profile fitting. For \CII~ and \SiIII, log$N$ is integrated over a certain velocity range with AOD method \citep{savage91, savage96}. For Component v-150, the integration range is $\vlsr=-150.2\pm(41.0/2)~\kms$, where $-150.2~\kms$ is the mean centroid velocity from \SiII, \SiIV, and \CIV, and $41.0~\kms$ is the mean FWHM ($\equiv 1.667b$). For Component v-220, the integration range is $\vlsr=-218.8\pm(58.0/2)~\kms$ defined in a consistent way. \HI\ column density is measured over the same velocity range as \CII\ and \SiIII\ using GBT spectrum from pointing 1 (see \S\ref{sec:gbt}). \textbf{Col (2)}: Centroid velocity for \CIV, \SiII, and \SiIV\ from Voigt-profile fitting, in LSR frame. \textbf{Col (3)}: Doppler width for \CIV, \SiII, and \SiIV\ from Voigt-profile fitting.  }
\begin{tabular}{lccc} 
\hline
Ion & log$N$          & $v_{\rm c}$  & $b$     \\
    & log (cm$^{-2}$) & (\mkms)      & (\mkms) \\
    & (1)             & (2)          & (3) \\
\hline
\hline
\multicolumn{4}{c}{Component v-150}\\
\hline
\HI    & $\leq18.02$ (3$\sigma$)  &  -    & -            \\
\CII   & $13.93\pm0.04$ & -               & -            \\
\CIV   & $13.67\pm0.09$ & $-156.0\pm4.8$  & $28.4\pm6.7$ \\
\SiII  & $12.97\pm0.07$ & $-151.7\pm2.4$  & $22.4\pm3.3$ \\  
\SiIII & $13.30\pm0.09$ & -               & -             \\
\SiIV  & $12.95\pm0.10$ & $-143.2\pm4.3$  & $23.0\pm6.6$ \\
\hline
\hline
\multicolumn{4}{c}{Component v-220}\\
\hline
\HI    & $\leq18.05$ (3$\sigma$)  &  -    & -            \\
\CII   & $14.07\pm0.03$ & -               & -            \\
\CIV   & $13.75\pm0.07$ & $-222.2\pm4.9$  & $32.7\pm6.2$ \\
\SiII  & $13.22\pm0.05$ & $-220.0\pm3.3$  & $40.2\pm5.8$ \\
\SiIII & $13.36\pm0.07$ & -               & -            \\
\SiIV  & $13.05\pm0.09$ & $-214.0\pm5.2$  & $31.5\pm8.0$ \\ 
\hline
\end{tabular}


\label{tb:ion}
\end{table}

We conduct continuum normalization and Voigt-profile fitting using the software developed for the COS-Halos survey \citep{tumlinson11, tumlinson13, werk13, werk14, prochaska17}. Details of the spectral analysis processes can be found in \cite{tumlinson13}; here we summarize the major procedures. Our analysis focuses on a number of ion absorption lines, including \SiII~1190/1193/1260/1526 \AA, \SiIII~1206 \AA, \SiIV~1394/1402 \AA, \CII~1334 \AA, and \CIV~1548/1550 \AA. Most of the ions have multiphase transition lines within the G130M/G160M spectral coverage, which solidify the line identification. Our data also cover \OI~1302 \AA, \PII~1152 \AA, \SII~1250/1253/1259 \AA, and \FeII~1142/1143/1144/1608 \AA, but do not show significant detection. The non-detection limit of these lines can be found in \citetalias{fox13} and \citeauthor{fox14} (2014; \citetalias{fox14} hereafter), and we do not use these lines in the following sections.

For each line of interest, we select a spectral region of $\pm1000~\kms$ within the rest wavelength for continuum normalization. We fit the absorption-line free regions with low-order Legendre polynomials until the reduced $\chi^2$ approaches 1.0. We proceed with Voigt-profile fitting by first visually inspecting the line profiles to estimate the number of velocity components and evaluate potential contamination from intercepting absorbers at higher redshifts. We run MPFIT \citep{markwardt09} to solve for the best-fit parameters of ion column density (log$N$), Doppler width ($b$), and velocity centroid ($v_c$). For ions with multiple lines, Voigt-profile fitting is conducted simultaneously for all the available transition lines to ensure consistent solutions. The line spread function of COS \citep{ghavamian09} has also been convolved into the fitting when we model the Voigt profiles.

We show the UV lines and Voigt-profile fitting results in Figure \ref{fig:cosspec}. Each of the transitions simultaneously present two velocity components at $v_{\rm LSR}\sim-150~\kms$ and $\sim-220~\kms$; hereafter, we refer to these two components as ``Component v-150" (color-coded in red) and ``Component v-220" (color-coded in blue) according to their LSR velocity measurements, respectively. For \CII\ and \SiIII, we do not attempt to decompose the line profiles because there is only one saturated transition for each ion.  In this case, the Voigt-profile fitting is highly uncertain. Instead, we define a spectral region of $-150.2\pm(41.0/2)~\kms$ equivalent to Component v-150. The region's center at $\vlsr=-150.2~\kms$ and its width of $41.0~\kms$ are from the mean $v_c$ and FWHM($\equiv1.667 b$) measured from the corresponding \SiII, \SiIV, and \CIV~ components. Similarly, we define a spectral region of $-218.7\pm(58.0/2)~\kms$ in \CII\ and \SiIII\ as Component v-220. We then estimate log$N$ for both components in \CII\ and \SiIII\ using the apparent optical depth method (AOD method; \citealt{savage91, savage96}). We tabulate the (log$N$, $b$, $v_c$) results in Table \ref{tb:ion}.

We note that \citetalias{fox13} and \citetalias{fox14} have studied the UV absorbers along PHL2525 in the context of Magellanic Stream (MS; see \S\ \ref{sec:origin}). They measure log$N$ for related absorbers with the AOD method over an integration range of $\vlsr=[-280, -120]~\kms$, which covers both Components v-150 and v-220. We compare their log$N$ estimates with the combined values of the two components listed in Table \ref{tb:ion}, and find that they are consistent within 0.11 dex despite the different methods and velocity ranges being adopted. Neither of us finds significant detection in \SII~ and \FeII. \citetalias{fox13} also report a potential detection of \OI\ 1302 \AA\ with an upper limit of log$N<14.30$ at $3\sigma$. Since it is a single line and only an upper limit is inferred from the data, we do not include this line in our analysis.

\subsection{GBT Observations and Ancillary Data}
\label{sec:gbt}
\begin{figure*}
\includegraphics[width=2\columnwidth]{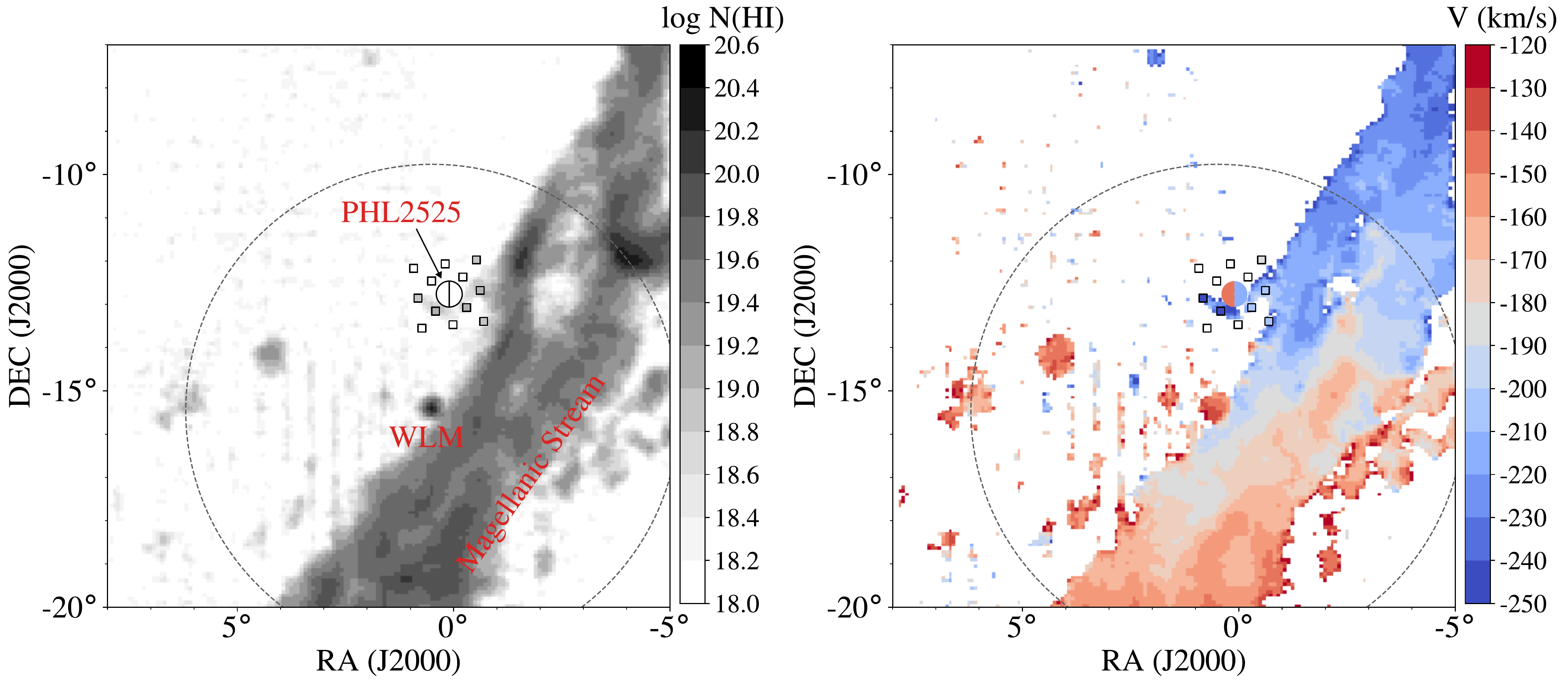}
\caption{Left: \HI\ column density ($N_{\rm HI}$) map of WLM and its nearby Magellanic Stream environment. The map is integrated from $\vlsr=-250~\kms$ to $-120~\kms$ using HI4PI data set \citep{hi4pi16}. The QSO PHL2525, at an impact parameter of 45.2 kpc ($0.5R_{\rm 200}^{\rm WLM}$), is noted with two wedges, representing the UV absorbers of Components v-150 and v-220 along the line of sight. 12 GBT pointings, as shown in squares, are observed surrounding PHL2525 in spacing of 30 arcmin. One additional pointing directly on the QSO is not shown for clear illustration. Right: flux-weighted velocity over the same velocity range. The colors of the wedges indicate \mvlsr\ values of Components v-150 and v-220. The GBT pointings are color-coded by the corresponding velocity when there is a detection; if none, we use white (see Figure \ref{fig:gbt} for GBT spectra). The pointing toward the QSO does not have a detection in HI emission (see \S\ \ref{sec:cos} for more details). The dotted circle in both maps shows the $R_{200}^{\rm WLM}$ radius of WLM. }
\label{fig:nhi}
\end{figure*}

To study the gaseous environment near PHL2525 and WLM, we generate \HI\ column density ($N_{\rm HI}$) and flux-weighted velocity maps using the HI4PI data set \citep{hi4pi16}, which is an all-sky \HI\ 21cm survey with angular resolution of $\theta_{\rm FWHM}=16.2$ arcmin. Assuming the \HI\ line is optically thin, the column density is calculated as $N$(\HI)$\equiv1.823\times10^{18}~[{\rm cm^{-2}~(K~\kms)^{-1}}]\int T(v) dv$ \citep{dickey90}, and the flux-weighted velocity as $\bar{v}\equiv \int vT(v)dv/\int T(v)dv$ (\mkms). We integrate the HI4PI data from $\vlsr=-250~\kms$ to $-120~\kms$ to include both the MS and part of the WLM emission\footnote{The \HI\ emission from WLM spreads over a range of $\vlsr\sim-170~\kms$ to $\sim-80~\kms$, see figure 7 in \cite{kepley07}.} as shown in Figure \ref{fig:nhi}. The noise level averaged over 1 \mkms\ is $\sigma_T\sim53$ mK (\citealt{hi4pi16}; also see table 1 in \citealt{peek18}); over a range of 130 \mkms, the $N_{\rm HI}$ map has a sensitivity of $\sigma_{\rm HI}=1.823\times10^{18}\times0.053*\sqrt{130}=1.1\times10^{18}$ cm$^{-2}$ at 1$\sigma$ (or $3.3\times10^{18}$ cm$^{-2}$ at $3\sigma$). Within one degree of PHL2525 (see below for GBT pointings), we find a peninsular-shape \HI~ feature extending from the main body of the Magellanic Stream with log$N_{\rm HI}\lesssim19.0$ (left panel) and $\vlsr\sim-250~\kms$ (right panel), which is close to the velocity of Component v-220 from UV spectra. We do not find any prominent \HI~ emission at $\vlsr\gtrsim-200~\kms$ that are consistent with Component v-150.

\begin{figure*}
\includegraphics[width=2\columnwidth]{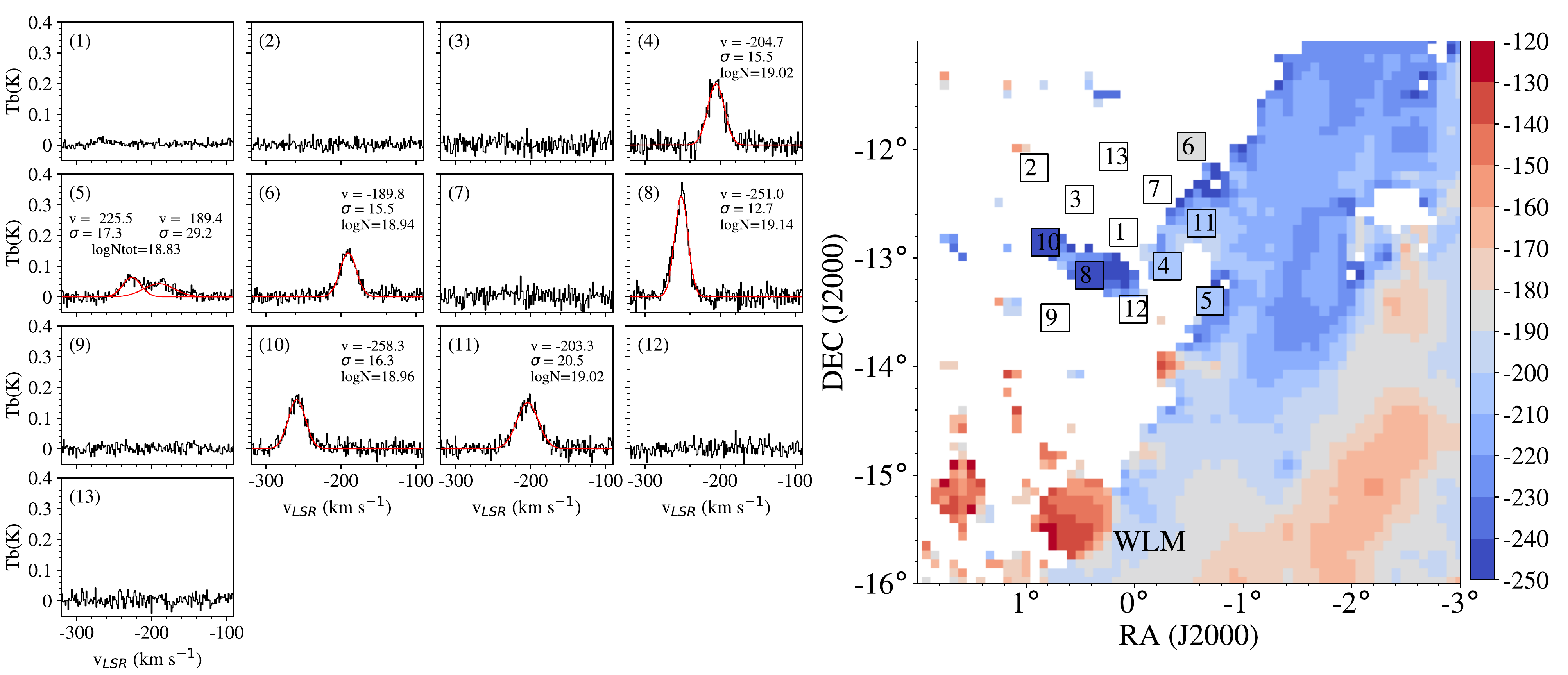}
\caption{\HI\ 21cm spectra observed with GBT along PHL2525 (pointing 1) and near the QSO in spacing of 30 arcmin (pointings 2-13). We fit pointing 4, 6, 8, 10, and 11 with single Gaussian and fit pointing 5 with double Gaussians. The fitted velocity $v$, velocity dispersion $\sigma$, and \HI\ column density log$N$ are also indicated accordingly; typical fitting errors are $\sim0.5~\kms$ for $v$ and $\sigma$, and 0.01 dex for log$N$. The right panel is a zoom-in version of the velocity panel in Figure \ref{fig:nhi}, with pointing number labelled consistently with the panel number on the left. In this figure, point 1 is directly toward PHL2525, which is not shown in Figure \ref{fig:nhi}. For pointing 4, 6, 8, 10, and 11, the symbols are color-coded by the velocities of Gaussian fits; for pointing 5, a mean velocity of the two Gaussians are adopted.}
\label{fig:gbt}
\end{figure*}

To examine the \HI\ structure near WLM and PHL2525 at a more sensitive level, we obtain 21cm spectra with the Robert C. Byrd Green Bank Telescope (GBT) of the Green Bank Observatory (proposal \#18B-376) which has a beam width (FWHM) of 9.1 arcmin. Observations were made by frequency-switching, and were calibrated and corrected for stray radiation as described in \cite{boothroyd11}. Final spectra cover -560 to +660 \mkms\ at a velocity resolution of 1.21 \mkms.  We observe sightlines toward WLM and PHL2525 directly, and use another 12 pointings on a coarse 30 arcmin grid to probe the \HI\ environment near the QSO sightline (see Figures \ref{fig:nhi} and \ref{fig:gbt}). Toward PHL2525 (pointing 1) the root-mean-square noise level is 8.1 mK, giving a $3\sigma$ detection limit for a 25 \mkms\ line of $2.2 \times 10^{17}$ cm$^{-2}$. The median noise level for the other directions (pointings 2-13) is 16 mK, giving a $3\sigma$ limit of $4.4 \times 10^{17}$  cm$^{-2}$. As shown in Figure \ref{fig:gbt}, there is significant \HI\ 21cm emission toward pointing 4, 5, 6, 8, 10, and 11. Gaussian fitting to these emission lines finds peak-flux velocities at $\sim[-190, -270]$ \mkms. The locations of GBT detection and the velocities are coincident with the peninsular-shape \HI\ feature shown in the $N_{\rm HI}$ map from HI4PI (see the right panel in Figure \ref{fig:gbt}).


For PHL2525 (pointing 1), we find an \HI\ upper limit of log$N\leq 18.05$ (3$\sigma$) for Component v-220 over $\vlsr=-218.2\pm(58.0/2)~\kms$, and log$N\leq18.02$ ($3\sigma$) for Component v-150 over $\vlsr=-150.2\pm(41.0/2)~\kms$ (see Table \ref{tb:ion} for explanation of the components' velocity ranges). 
A weak feature may exist near $\vlsr\sim-260~\kms$ with log$N\sim18.23$ over a velocity range of 30 \mkms, consistent with the one detected in \citetalias{fox13} using the LAB data set \citep{Kalberla05}. \citetalias{fox13} relates this feature to the UV absorption of Components v-150 and v-220 despite a velocity offset of $\gtrsim50~\kms$. They suggest that the velocity mismatch is due to the large beam size (36 arcmin) of the LAB data. However, our higher-resolution HI4PI/GBT data suggest that the beam size is unlikely to be the culprit. In the next section, we argue that gas with different origins may result in such velocity mismatch.

\section{Origins of UV Absorbers}
\label{sec:origin}

PHL2525 is $45.2$ kpc from WLM in projection (see Figure \ref{fig:nhi}). As described in \S\ \ref{sec:intro}, we adopt the virial radius of WLM as $R_{\rm 200}^{\rm WLM}$=89.7kpc; therefore, PHL2525 is at 0.5$R_{\rm 200}^{\rm WLM}$ in WLM's CGM. Alternatively, the sightline passes through the edge of the Magellanic Stream (MS), which is a one-hundred-degree long \HI\ structure originating from the Large and Small Magellanic Clouds \citep{putman03, nidever08}. Due to this proximity, it is possible that the two UV components originate either from MS-related gas or from ionized gas in the CGM of WLM. Here we explore these two possible origins separately.

\begin{figure}
\includegraphics[width=\columnwidth]{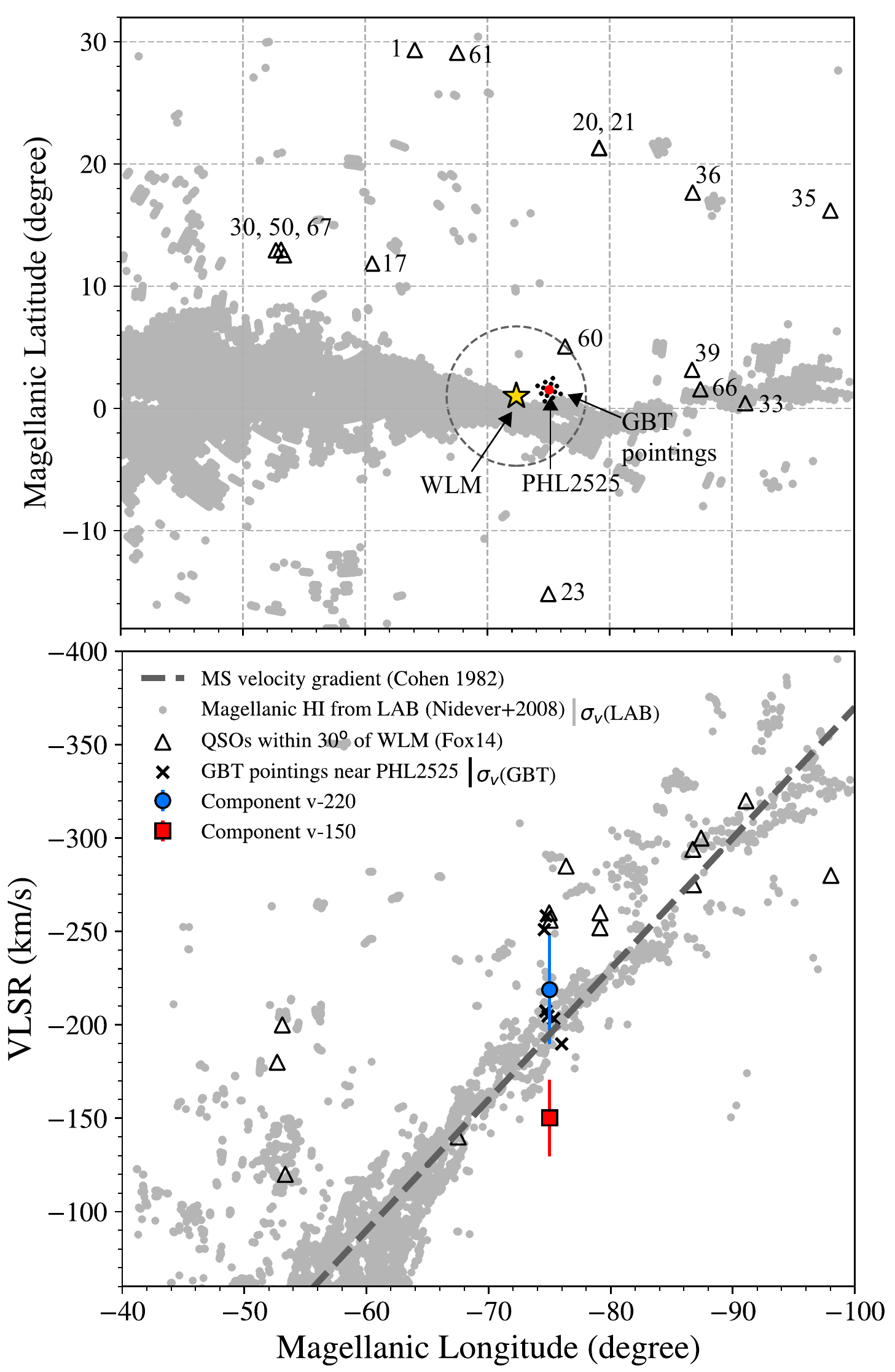}
\caption{Position-velocity diagram for \HI\ and ionized gas near WLM. In both panels, grey dots are \HI\ 21cm measurements of the MS \citep{nidever08}, and black open triangles are ion absorbers detected along QSO sightlines within 30 degrees of WLM by \citetalias{fox14}. In the top panel, we plot WLM as a yellow star, PHL2525 as a red dot, and the GBT pointings near PHL2525 as smaller black dots. The large dashed circle indicates the virial radius of WLM. The numbers near the QSOs are target IDs assigned in \citetalias{fox14}. PHL2525 is ID 41 in their work. In the bottom panel, we show the \HI\ detection from the GBT pointings (see Figure \ref{fig:gbt}) as black crosses. The blue dot is for Component v-220 and the red square for Component v-150, both with error bars from the Voigt-profile fitting (see Table \ref{tb:ion}). For clear illustration, we indicate the median error for the GBT HI detection ($\sigma_v=16.3~\kms$) as a black vertical line and that of the LAB \HI\ detection ($\sigma_v=14.8~\kms$) as a grey line in the figure legend. No error bars are given for the ion absorbers (triangles) in \citetalias{fox14}. We also plot the MS velocity gradient ($dv/dl_{\rm MS}\sim7~\kms~{\rm deg}^{-1}$; \citealt{cohen82, putman03, nidever10}) as a thick long-dashed line. }
\label{fig:pvplot}
\end{figure}

\subsection{Magellanic Stream Origin}

In fact, \citetalias{fox13} and \citetalias{fox14} included PHL2525 in their UV absorption survey to study MS's ionized extension, broadly defined as areas within 30 degrees of \HI-bright regions of the MS (see figure 1 in \citetalias{fox14}). In their study, a spectral region of $-280\leq\vlsr\leq-120~\kms$ is assumed for MS-related gas along PHL2525, covering both Components v-150 and v-220. Here we re-evaluate the relation of both components to the MS focusing on their position-velocity alignment. We do not rely on a metallicity estimate given the lack of significant \HI\ detection which would introduce a large uncertainty due to an unknown total hydrogen content.

In Figure \ref{fig:pvplot} we compare the positions and velocities of the two components with nearby \HI\ emission and QSO absorption line measurements. \cite{nidever08} conducted Gaussian decomposition of \HI\ emission lines from the MS observed with the LAB survey (Kalberla et al. 2005). The Magellanic longitudes, latitudes\footnote{The Magellanic longitude $L_{\rm MS}$ and latitude $B_{\rm MS}$ are defined by \cite{nidever08} as the Magellanic Stream coordinate system. In this system, the LMC is at $L_{\rm MS}=0$ and the MS extends from $L_{\rm MS}=0$ to $L_{\rm MS}<-100$ degree. Gas to the north of the MS is at $B_{\rm MS}>0$ degree and those to the south is at $B_{\rm MS}<0$ degree. }, and the Gaussian-fitted centroid velocities are plotted as grey dots in the figure. We search for QSOs within 30 degrees of WLM and find 15 sightlines in addition to PHL2525 from \citetalias{fox14}; these 15 sightlines are shown as black open triangles. \citetalias{fox14} analyzed the HST/COS archival spectra for these sightlines and determined MS-related centroid velocities based on either \HI\ emission (if any) or the strongest absorption components detected along the lines of sight. As shown in the bottom panel, near the Magellanic longitude of PHL2525 ($l_{\rm MS}$=75$^{\rm o}$), the MS-related absorbers all lie either on the MS main body or in regions with more negative velocities ($\lesssim$-190~\mkms) where \HI\ debris is scattered. Moreover, the \HI\ emission detected near PHL2525 through the GBT pointings also show velocities consistent with the MS. No \HI\ emission is detected at the velocity of Component v-150.

We find that the location of Component v-220 (blue circle) in the bottom panel coincides with the Magellanic \HI\ emission and UV absorption. However, Component v-150  (red square) is substantially below the Magellanic main body, showing an opposite trend from the ion absorbers detected in other nearby QSOs, which are located \emph{above} the main body of the Stream in the p-v diagram. Based on the position-velocity alignment, Component v-220 is most likely to originate from the MS while Component v-150 arises from a non-Magellanic origin.

Line ratio diagnostics are often used to study physical properties of ionized gas \citep[e.g.][]{fox11, wakker12, werk16}. We find that Component v-150 has line ratios of log($N_{\rm SiIV}/N_{\rm SiII}$)$=-0.02\pm0.12$, log($N_{\rm SiIII}/N_{\rm SiII}$)$=0.33\pm0.11$, and log($N_{\rm CIV}/N_{\rm CII}$)$=-0.26\pm0.10$ dex, whereas those of Component V-220 are $-0.17\pm0.10$, $0.14\pm0.09$, and $-0.32\pm0.08$ dex, respectively. The line ratios of Component v-150 is marginally higher than those of Component v-220 by $0.1-0.2$ dex ($\sim1-2\sigma$). For the MS gas, \citetalias{fox14} measure the line ratios as a function of Magellanic longitude $l_{\rm MS}$ \citep{nidever08}. PHL2525 is at $l_{\rm MS}=-75.0$ degree. They report a broad line ratio range for the MS gas at a similar longitude as PHL2525: log$(N_{\rm SiIV}/N_{\rm SiII})\sim[-0.5, 0.2]$, log$(N_{\rm SiIII}/N_{\rm SiII})\sim[-0.6, 0.4]$, and log$(N_{\rm CIV}/N_{\rm CII})\sim[-0.9, 0.4]$ (see figure 4 and table 2 in \citetalias{fox14}). The line ratios for both of Components v-150 and v-220 are within the quoted Magellanic range. Given the broad Magellanic range and the uncertainties involved, the line ratios are inconclusive with regard to membership identification.




Overall, the position-velocity analysis of nearby Magellanic \HI\ emission and ion absorbers indicate a Magellanic velocity range that accounts for Component v-220, but not Component v-150. Furthermore, Component v-220 is $\delta v\sim90~\kms$ from WLM's systemic velocity, which is much higher than the escape velocity of the galaxy's halo ($v_{\rm esc}\sim50~\kms$). Therefore, it is unlikely that Component v-220 resides in the CGM of WLM.

\subsection{WLM's CGM Origin}
The centroid velocity of Component v-150 differs from that of Component v-220 by $\delta_v\sim70~\kms$. Such a high velocity difference is unlikely to be caused by the velocity scatter of shredded MS debris clouds as we have seen from the UV and \HI\ data. Because PHL2525 is at 0.5$R_{\rm 200}^{\rm WLM}$ from WLM, here we explore the possibility that Component v-150 is a detection of the CGM of the galaxy. WLM moves at $\vlsr\sim-132~\kms$; in the galaxy's reference frame, Component v-150 moves at $\sim20~\kms$, which is faster than the escape velocity of the galaxy ($v_{\rm esc}\sim17~\kms$) but slower than that of the dark matter halo ($v_{\rm esc}\sim50~\kms$). Therefore, Component v-150 can be well retained in the CGM of WLM. 

SFH analysis of WLM has shown that the galaxy experienced an active star forming phase $1-3$ Gyrs ago \citep{dolphin00, weisz14a}. Supposing the metals ejected due to stellar feedback during this phase travel into the CGM at $\sim20~\kms$ and there is no significant velocity loss, we would find the gas at $R\sim20-60$ kpc at the current time, bracketing the location of Component v-150 in WLM's CGM. Additionally, we find log$N_{\rm CIV}=13.67$ for this absorber (see Table \ref{tb:ion}). This value is consistent with the estimate by the COS-Dwarf survey \citep{bordoloi14} which detect \CIV\ absorption out to $\sim0.5 R_{\rm vir}$ at log$N_{\rm CIV}\sim13.7$, although we note that WLM has a lower stellar mass than those in the dwarf sample studied by \cite{bordoloi14}.

Lastly, it is unlikely that both or either of the components are related to the CGM of the Milky Way (MW) and lie in the foreground of the MS. As shown in Figure \ref{fig:cosspec}, the absorption caused by the MW's CGM can be found at $\vlsr\gtrsim-100~\kms$, and no apparent HI high-velocity cloud exists along the direction of PHL2525 besides the Magellanic Stream at $\vlsr\lesssim-200~\kms$, as shown in the sensitive GBT pointings in Figure \ref{fig:gbt}. 


\section{Discussion: WLM's Silicon Budget}
\label{sec:discuss}

In this section, we first focus on the implication of Component v-150 in the context of WLM's CGM (\S \ref{sec:cgmmass}), then estimate the silicon (Si) mass budget in stars, ISM, CGM, and IGM accordingly in \S\ \ref{sec:massbudget}. 

\subsection{Silicon in WLM's CGM}
\label{sec:cgmmass}
Because multiple silicon and carbon ions are simultaneously detected whereas there is no trace of \HI~ at a sensitivity of log$N_{\rm HI}\gtrsim18.02$ at $3\sigma$, the CGM of WLM should be well ionized and metal-enriched. Assuming that Si is mostly in the forms of \SiII, \SiIII, and \SiIV, to the zeroth order, we can solve for the total Si mass contained in WLM's CGM as the following. From Table \ref{tb:ion}, we find that the total Si column density at $0.5R_{\rm 200}^{\rm vir}$ is log$N_{\rm Si}\equiv$log$N_{\rm SiII+SiIII+SiIV}=13.6$ as measured from Component v-150. Assuming a constant density profile in WLM's CGM, we have:
 \begin{equation}
 \begin{split}
     M_{\rm Si, 0th}^{\rm CGM} & = \pi (R_{200}^{\rm WLM})^2 m_{\rm Si}N_{\rm Si} C_{\rm f}\\
     & \approx 2\times10^4~\msun (\frac{N_{\rm Si}}{10^{13.6}~{\rm cm^{-2}}}) (\frac{R_{200}^{\rm WLM}}{89.6~{\rm kpc}})^2 (\frac{C_{\rm f}}{0.4}), 
\end{split}
\label{eqn:msi_0th}
 \end{equation}
where $C_{\rm f}$ is the covering fraction (i.e., detection rate). Because we only have one sightline through the halo, adopting a $C_f$ value of unity is unpractical. There are currently no observational constraints on the $C_{\rm f}$ value of Si in WLM's mass range. The best comparison point is the \CIV\ survey by the COS-dwarfs team \citep{bordoloi14} which covers dwarf mass range of $M_*\sim10^{8.2-10.2}~\msun$. We assume that their results can be extrapolated to WLM's mass and that \SiII, \SiIII, \SiIV, and \CIV\ co-exist in a multiphase medium. We find  $C_{\rm f}\approx0.4$\footnote{17 of the 43 QSO sightlines from the COS-Dwarf sample \citep{bordoloi14} show detection or lower limits of \CIV. }. 



\begin{table*}
\centering
\caption{\textbf{Silicon mass estimates for WLM}. }
\begin{tabular}{clcl} 
\hline
 & Name  & Mass (\mmsun)      & Note     \\
\hline
\hline
(1) & $M_*$ & $4.3\times10^7$ &  Stellar mass \citep{jackson07} \\
\hline
(2) & $M_{\rm HI}$ & $(6.3\pm0.3)\times10^7$ & \HI\ mass \citep{kepley07}; total gas mass in the ISM including He is $\sim8.5\times10^7~\msun$  \\
\hline
(3) & $M_{\rm h}{\rm (ISO)}$ & $(2.6\pm0.2)\times10^{10}$ & Halo mass assuming isothermal spherical volume \citep{leaman12} \\
\hline
\multirow{2}{*}{(4)} & \multirow{2}{*}{$M_{\rm Si}^{\rm CGM}$} & \multirow{2}{*}{$\sim(0.2-1.0)\times10^5$} & Estimate of the CGM Si mass based on QSO absorption line data,   \\
 & & & the range is given based on the zero-th order and the refined calculations in Eq. \ref{eqn:msi_0th} and \ref{eqn:msi_refine}\\
 \hline
\multirow{3}{*}{(5)} & \multirow{3}{*}{$M_{\rm SF}^{\rm tot}$} & \multirow{3}{*}{$\sim6.5\times10^7$} & Total stellar mass formed throughout WLM's SFH, $M_{\rm SF}^{\rm tot}$=$M_*/(1-R)$, see Eqn. \ref{eqn:mtotsf}. \\
    & & &$R=0.34$ is the fraction of mass returned to ISM from a stellar population, and $R$ can vary \\
    & & &by $\sim35\%$ with different choice of IMFs and by $10-30\%$ with different stellar yield sets.\\
\hline
\multirow{2}{*}{(6)} & \multirow{2}{*}{$M_{\rm Si}^{\rm gas}$} & \multirow{2}{*}{$\sim1.3\times10^5$} & Expected total Si mass in gas forms in ISM, CGM, and IGM, see Eqn. \ref{eqn:msigas}. $y_{\rm Si}=0.003$ is the \\
 & & &net stellar yield, and its value can vary by $\sim30\%$ due to different choice of metallicity.\\
 \hline
\multirow{2}{*}{(7)} & \multirow{2}{*}{$M_{\rm Si}^{\rm ISM}$} & $(7.9\pm1.5)\times10^3$ & Total Si mass in ISM, based on $M_{\rm HI}$ and oxygen abundance of WLM's \HII\ regions, see Eqn. \ref{eqn:msiism}. \\
 & & & $M_{\rm Si}^{\rm ISM}$ would be a factor of two higher if WLM had a similar Si depletion pattern as SMC. \\
\hline
(8) & $M_{\rm Si}^{\rm *}$ & $(4.0\pm0.7)\times10^3$ & Total Si mass in stars, assuming the stars have similar composition as the ISM, see Eqn. \ref{eqn:msi*}. \\
\hline
(9) & $M_{\rm Si}^{\rm tot}$ & $\sim1.3\times10^5$ & Total Si mass that have been produced, $M_{\rm Si}^{\rm tot}$=$M_{\rm Si}^{\rm gas}$+$M_{\rm Si}^*$ \\
\hline
\hline
\end{tabular}
\label{tb:mass}
\end{table*}

Assuming a constant density profile in Equation \ref{eqn:msi_0th} may not best represent the nature of the gas distribution in WLM's CGM. As pointed out by \cite{oh15}, the mass distribution of dwarf galaxies observed in the LITTLE THINGS survey \citep{hunter12} tends to have a smoother isothermal profile than an NFW profile. With the large uncertainty of the  matter density distribution in dwarf galaxies in mind, we can refine the $M_{\rm Si, 0th}^{\rm CGM}$ value by calculating the mass over the whole halo in an annular manner; the method is also commonly adopted to estimate the CGM mass for extragalactic systems (e.g., \citealt{werk14, peeples14, lehner15, prochaska17}). Taking $N_{\rm Si}(r)$ as the Si column density profile and $C_{\rm f}(r)$ the covering fraction profile, we have: 
\begin{equation}
    M_{\rm Si, refined}^{\rm CGM}=\int_0^{R_{\rm 200}^{\rm WLM}} m_{\rm Si} N_{\rm Si}(r)C_{\rm f}(r)2\pi r {\rm d}r\sim1\times10^5~\msun. 
\label{eqn:msi_refine}
\end{equation}
In this calculation, we again use the \CIV\ measurements from the COS-Dwarfs survey \citep{bordoloi14} as a proxy to derive $N_{\rm Si}(r)$ and $C_{\rm f}(r)$ profile assuming a multiphase medium. Using all of the detections found in their table 1 (and treating lower limits as detection as well), we find a power-law distribution of $N_{\rm CIV}(r)\propto (\frac{r}{R_{\rm vir}})^{-1.6}$. Normalizing this profile at $r/R_{\rm vir}=0.5$ with logN$_{\rm Si}=13.6$ for WLM's CGM, we have $N_{\rm Si}(r)=10^{13.1}(\frac{r}{R_{\rm vir}})^{-1.6}$. Similarly, for the covering fraction profile, we find, $C_{\rm f}(r/R_{\rm vir}\leq0.2)=0.9$, 
$C_{\rm f}(0.2<r/R_{\rm vir}\leq0.4)=0.5$, and $C_{\rm f}(0.4<r/R_{\rm vir}\leq0.6)=0.2$ from their data; beyond 0.6$R_{\rm vir}$, the \CIV\ detection rate drops to zero. Because we only aim for a coarse estimate, therefore, we do not take into account the detection limit when estimating $C_{\rm f}(r)$ from their data. We note that both $M_{\rm Si, 0th}^{\rm CGM}$ and $M_{\rm Si, refined}^{\rm CGM}$ are only as accurate as order-of-magnitude estimates given the assumptions that go into the calculations. The major sources of uncertainties in our calculation are from the virial radius of the galaxy's dark matter halo, the Si density profile, and the covering fraction. 
In the following, we adopt $M_{\rm Si}^{\rm CGM}=(0.2-1)\times10^5~\msun$ which incorporates both of the calculations in Eqn. \ref{eqn:msi_0th} and \ref{eqn:msi_refine}. In Table \ref{tb:mass}, we record the $M_{\rm Si}^{\rm CGM}$ value as well as other mass estimates described as follows.

\subsection{Silicon in Stars and ISM}
\label{sec:massbudget}
We can compare $M_{\rm Si}^{\rm CGM}$ with the total amount of Si that has ever been produced in WLM using stellar evolution models and star-formation history analysis. A similar technique is also used by \cite{telford18} to estimate the metal loss from M31. Assuming a Kroupa initial mass function (IMF; \citealt{kroupa02}) with a minimum stellar mass of $0.08~\msun$ and a maximum of $100~\msun$ at the metallicity of WLM, we find that the fraction of mass returned to the ISM per stellar generation is $R=0.34$. The net stellar yield of Si is $y_{\rm Si}\sim0.003$, which is defined as the ratio of Si mass produced and available in gas to the amount of mass locked in stars, $y_{\rm Si}=M_{\rm Si}^{\rm gas}/M_{\rm *}$ (also see equation 2, \citealt{vincenzo16}). The values of $R$ and $f_{\rm Si}$ are calculated using the NuGrid collaboration yield set \citep{ritter18b} and the SYGMA simple stellar population model \citep{ritter18a}. Detailed calculations with different choices of IMF, stellar mass range, and metallicity are shown in Appendix \ref{sec:app} and Table \ref{tb:stellaryield}. We find that $R$ is not sensitive to the choice of metallicity -- it only varies by $\sim2$\%\ from $Z=0.0001$ to 0.02 $Z_\odot$. However, choosing a Salpeter IMF \citep{salpeter1955} would reduce the value by $\sim35\%$ from $R\approx0.34$ to $R\approx0.23$. On the other hand, the stellar yield $y_{\rm Si}$ does not vary significantly with different IMFs, but is sensitive to the choice of metallicity. We expect a variation of $\sim30\%$ from $Z=0.0001$ to 0.02 $Z_\odot$. We also consider the influence of different yield tables by comparing our $R$ values with those in \cite{vincenzo16}. We find that, at a fixed IMF and metallicity, $R$ varies by $\sim10-30\%$ with different stellar yield sets from \cite{romano10}, \cite{nomoto13}, and \cite{ritter18b}. In the following calculation, we proceed with $R=0.34$ and $y_{\rm Si}=0.003$, and note that these values are subject to the details of different models of stellar evolution and stellar yields. 

The fraction of mass locked in stars is $R_*=1-R=0.66$.  
Since WLM has a stellar mass of $M_*=4.3\times10^7~\msun$ \citep{jackson07, mcconnachie12}, in total the galaxy has formed 
\begin{equation}
M_{\rm tot}^ {\rm SF}=M_*/(1-R)\sim6.5\times10^7~\msun
\label{eqn:mtotsf}
\end{equation}
throughout its star-formation history. This mass estimate is consistent with the SFH analysis on combined UVIS and ACS fields in WLM by \cite{albers19}, who show that a total amount of $M_{\rm UVIS+ACS}\approx7.4\times10^6~\msun$ has been formed over $\sim50$\% of the area within the half-light radius ($r_{\rm h}$= 2.1 kpc) of the galaxy (see also \citealt{weisz14a}). Assuming a constant mass-to-light ratio over the whole galaxy, we find a total mass of $M_{\rm UVIS+ACS}/0.5\times2\sim3\times10^7~\msun$ formed within twice the half-light radius. This value is consistent with the estimated $M_{\rm tot}^{\rm SF}$ value in Eq. \ref{eqn:mtotsf} given the uncertainty in $R$. Using the net stellar yield $y_{\rm Si}$, we find that the total Si mass now in the form of gas is 
\begin{equation}
M_{\rm Si}^{\rm gas}=y_{\rm Si}M_*\sim1.3\times10^5~\msun
\label{eqn:msigas}
\end{equation}
, which is expected to be in WLM's ISM, CGM, and the surrounding IGM. Note that this calculation assumes instantaneous and homogeneous mixing, which may slightly underestimate $M_{\rm Si}^{\rm gas}$.

Independently, we can estimate the total Si mass in WLM's ISM, $M_{\rm Si}^{\rm ISM}$, as the following. A similar technique is used by \cite{mcquinn15} to estimate the oxygen mass in the ISM and stars of Leo P (see their Eqn. 2 and 4). WLM has an oxygen abundance of 12+log(O/H)$_{\rm WLM}$
$=7.83\pm0.06$ as measured from its \HII\ regions \citep{lee05}. Assuming that WLM's ISM has a similar element composition as the Sun, despite its lower metallicity, we can estimate the Si abundance as 
\begin{equation}
\begin{array}{ll}
12+{\rm log(Si/H)_{WLM}}&={\rm log[(\frac{Si}{O})(\frac{O}{H})]}\\
&={\rm log(Si/O)_\odot+log(O/H)_{WLM}}\\
&=6.65\pm0.08
\end{array}
\end{equation}
, where log(Si/O)$_\odot\equiv$log(Si/H)$_\odot-$log(O/H)$_\odot$, and we adopt Si and O solar abundance 12+log(Si/H)$_\odot=7.51\pm0.03$ and 12+log(O/H)$_\odot=8.69\pm0.05$ from \cite{asplund09}. Assuming that molecular gas contributes little to WLM's ISM gas mass \citep{rubio15}, the total Si mass in the ISM is:
\begin{equation}
M_{\rm Si}^{\rm ISM}=M_{\rm HI}(\frac{m_{\rm Si}}{m_{\rm H}})10^{\rm log(Si/H)_{WLM}}=(7.9\pm1.5)\times10^3 \msun
\label{eqn:msiism}
\end{equation}
, where $m_{\rm Si}$ and $m_{\rm H}$ are the Si and H atomic mass numbers, respectively. In the above calculation we assume that the amount of Si depleted in dust is negligible in WLM's ISM. This calculation may underestimate the amount of Si in the ISM by a factor of two if WLM followed a similar Si depletion pattern as the Small Magellanic Cloud (SMC) which has a similar metallicity as WLM. \cite{Jenkins17} study the gas-phase abundance and element depletion of SMC's ISM using UV absorption lines of 18 stars in the galaxy, and find that that amount of Si in dust with respect to the amount of total H is ${\rm (Si_{dust}/H)_{SMC}}\approx4\times10^{-6}$ (see their Eqn. 1, 5 and table 3). If applicable to WLM, we would find a total mass of $M_{\rm Si}^{\rm dust}=(m_{\rm Si}/m_{\rm H})M_{\rm HI}{\rm (Si_{dust}/H)_{SMC}}\approx7\times10^3~\msun$ in dust. In such case, the $M_{\rm Si}^{\rm ISM}$ value calculated in Eqn. \ref{eqn:msiism} should be a factor of two higher.

Similarly, if we assume that the stars share similar composition with the ISM, the total amount of Si locked in the stars can be calculated as:
\begin{equation}
\begin{array}{l l}
M_{\rm Si}^*&=M_*\frac{\rho_{\rm Si}}{\rho_{\rm H+He}}=M_*\frac{m_{\rm Si}n_{\rm Si}}{(m_{\rm H}n_{\rm H}/X)}=M_*(\frac{m_{\rm Si}}{m_{\rm H}})(\frac{n_{\rm Si}}{n_{\rm H}})X\\
      &=(4.0\pm0.7)\times10^3~\msun, 
\end{array}
\label{eqn:msi*}
\end{equation}
, where $X=0.74$ is the hydrogen mass fraction and $(\frac{n_{\rm Si}}{n_{\rm H}})=10^{\rm log(Si/H)_{WLM}}$.

In all, the total amount of Si that has been produced is $M_{\rm Si}^{\rm tot}=M_{\rm Si}^{\rm gas}+M_{\rm Si}^*\sim1.3\times10^5~\msun$. Of this mass, $\sim3\%$($=M_{\rm Si}^{\rm *}/M_{\rm Si}^{\rm tot}$) is locked in stars, $\sim6\%$($=M_{\rm Si}^{\rm ISM}/M_{\rm si}^{\rm tot}$) is retained in the ISM, and $\sim15\%-77\%$($=M_{\rm Si}^{\rm CGM}/M_{\rm Si}^{\rm tot}$) is in the CGM. The remainder Si, $14\%-76\%$, may be depleted in dust in the ISM, exist in higher ionization phases in the CGM, or have been blown out into the IGM. Dust depletion may account for another 6\% of the silicon as we discuss near Eqn. \ref{eqn:msiism}, whereas the amount of Si at higher ionization states should be minimal. Recent simulations have shown that $\sim90\%$ of the metals in the CGM of low-mass dwarf galaxies are in cool ($T\sim10^{4-4.7}$ K) phase that we have already probed in \SiII, \SiIII, and \SiIV\ \citep{muratov17}, as compared to a much lower fraction ($19-42\%$) in cool phase in $\sim L*$ galaxies \citep{oppenheimer18}.

To summarize, our calculation is consistent with existing observations of both Leo P \citep{mcquinn15} and dwarf spheroidal galaxies of the Milky Way \citep{kirby11} which estimate that $\gtrsim95\%$ of the metals synthesized by stars in dwarf galaxies have been lost into the CGM and IGM. It also supports theoretical predictions that metals are easily lost to the CGM in low-mass galaxies through feedback-driven outflows \citep[e.g.][]{maclow99, ma16, muratov17, christensen18}. It is beyond the scope of this work to explore further the detailed mass fraction differences between our calculation and the theoretical predictions. We note that it is challenging to determine metal mass distribution conclusively in part because the number of detailed simulations of dwarf galaxies in this mass range is limited, and the properties of simulated dwarfs at this stellar mass for a single feedback prescription exhibit noticeable variability. In addition, our CGM mass calculation is subject to observational and model-dependent uncertainties, such as the $M_*-M_{\rm h}$ relation, assumed halo density profile, assumed covering fraction, and the limited studies of the CGM in dwarf galaxies at such a low-mass ($M_*\sim10^7~\msun$) regime. Therefore, the reader should use caution when interpreting results from both simulations and our observation.


\section{Summary}
\label{sec:sum}

We analyze an archival HST/COS spectrum of a QSO sightline, PHL2525, passing through the CGM of WLM at an impact parameter of 45.2 kpc ($R_{\rm 200}^{\rm WLM}$). Two absorption line components are recovered with Voigt-profile fitting at $\vlsr\sim-220$ and $\sim-150~\kms$ (i.e., Components v-220 and v-150). Both WLM and PHL2525 are in the vicinity of the MS in projection. we investigate the relations of Components v-220 and v-150 with the ionized extension of MS and the CGM of WLM. Our findings are summarized as the following.

First, based on the position-velocity diagram in Figure \ref{fig:pvplot}, we find that at the Magellanic longitude around $l_{\rm MS}=-75$ degree, both the neutral (\HI) and ionized gas related to the MS are moving at velocities of $\vlsr\lesssim-190~\kms$. The velocity of Component v-220 is consistent with that of the MS, whereas Component v-150 is beyond this range. Because PHL2525 is at 0.5 $R_{\rm 200}^{\rm WLM}$ from WLM and Component v-150 moves at $\sim20~\kms$ within respect to the galaxy's systemic velocity, we propose that Component v-150 arises from the CGM of WLM. SFH analysis of WLM shows that the galaxy experienced an active star-forming epoch 1-3 Gyrs ago. If WLM expelled metals into the CGM during this epoch, the gas parcel would have travel to a distance of $20-60$ kpc assuming a speed of $20~\kms$, which is consistent with the location of Component v-150 at $R=45.2$ kpc from WLM.

If Component v-150 is associated with the CGM of WLM, we find a total Si mass of $M_{\rm Si}^{\rm CGM}\sim(0.2-1.0)\times10^5~\msun$ within the virial radius of the galaxy. This assumes all the silicon atoms are in the forms of \SiII, \SiIII, and \SiIV. We note that the $M_{\rm Si}^{\rm CGM}$ value should be treated only as an order-of-magnitude approximation with the assumptions involved. We also calculate the Si mass in the stars and ISM using stellar evolution modeling, and find $M_{\rm Si}^*=(4.0\pm0.7)\times10^3~\msun$ in stars and $M_{\rm Si}^{\rm ISM}=(7.9\pm1.5)\times10^3~\msun$ in the ISM (see Table \ref{tb:mass}). Therefore, the Si mass fractions now in the forms of stars, ISM and CGM of WLM are $\sim3\%$, $\sim$6\%, and $\sim15-77$\%, respectively. The rest of the Si, $14-76\%$, may be depleted in the dust in the ISM, exist in higher ionization phases other than \SiII, \SiIII, and \SiIV\ in the CGM, or have been blown out into the IGM. Our finding is consistent with theoretical predictions that metals can be easily expelled to the CGM in low-mass galaxies.

Because WLM is isolated from other galaxies in the LG, the detection of its CGM as revealed by Component v-150 provides an unique case to study how stellar feedback distributes metals into the CGM of low-mass dwarf galaxies without environmental influence. It supports theoretical work (e.g., \citealt{maclow99, ma16, muratov17, christensen18}) which predict that dwarf galaxies with $M_*\sim10^{7-8}~\msun$ cannot retain most of the metals they have synthesized, and that most of the lost metals would reside in the CGM at cool phases. To better understand how metals are transported through stellar feedback processes, it is of great need to conduct further observations of the CGM of low-mass dwarf galaxies in the $M_*\lesssim 10^8~\msun$ regime, and to include SFH analyses of the galaxies' metal production history.

\section*{Acknowledgements}
Y.Z. is grateful to D. Weisz for many useful discussions during the preparation of this work. Y.Z. thanks S. Albers for providing SFH estimates of WLM on combined UVIS/ACS fields, and thanks C. F. McKee for his feedback on this manuscript.
Y.Z. acknowledges support from the Miller Institute for Basic Research in Science. 
A.E. was supported by the National Science Foundation (NSF)
Blue Waters Graduate Fellowship. 
E. N. K. acknowledges support by NASA through a grant (HST-GO-15156.004-A) from the Space Telescope Science Institute, which is operated by the Association of Universities for Research in Astronomy, Incorporated, under NASA contract NAS 5-26555. Facilities: HST/COS, Green Bank Telescope. Software: Astropy \citep{astropy13}, IDL, Numpy \citep{numpy}, Matplotlib \citep{hunter07}




\bibliographystyle{mnras}
\bibliography{main} 



\appendix

\section{Stellar Evolution Modeling}
\label{sec:app}
We calculate the fraction of mass returned into the ISM per stellar generation ($R$) and the net stellar yield ($y_{\rm Si}$) using the NuGrid collaboration yield set \citep{ritter18b} and the SYGMA simple stellar population model \citep{ritter18a}. The definitions of $R$ and $y_{\rm Si}$ follow those in \cite{vincenzo16}. In Table \ref{tb:stellaryield}, we show the $R$ and $y_{\rm Si}$ values calculated with different choices of IMFs and metal mass fractions. For the calculation of each set of ($Z$, $R$, $y_{\rm Si}$), we assume a minimum stellar mass of $0.08~\msun$ and a maximum of $100~\msun$. When compared with the results from \cite{vincenzo16}, the value of $R$ varies by $\sim10-30\%$ for a given IMF and $Z$, probably due to different choices of stellar yield sets and a different treatment of massive stars in the stellar evolution modeling. 

\begin{table}
\centering
\caption{\textbf{Stellar Yields $y_{\rm Si}$ and Gas Return Fraction $R$}. }
\begin{tabular}{ccc} 
\hline
$Z$ & $R$          & $y_{\rm Si}$  \\
\hline
\hline
\multicolumn{3}{c}{Kroupa IMF \citep{kroupa02} }\\
\hline
0.0001 & 0.343 & 0.00269  \\
0.0010 & 0.342 & 0.00311  \\
0.0060 & 0.347 & 0.00372  \\
0.0100 & 0.349 & 0.00333  \\  
0.0200 & 0.349 & 0.00334  \\
\hline
\hline
\multicolumn{3}{c}{Salpeter IMF \citep{salpeter1955} }\\
\hline
0.0001 & 0.228 & 0.00281 \\
0.0010 & 0.227 & 0.00320 \\
0.0060 & 0.230 & 0.0376  \\
0.0100 & 0.232 & 0.00342 \\
0.0200 & 0.232 & 0.00345 \\
\hline
\end{tabular}
\label{tb:stellaryield}
\end{table}



\bsp	
\label{lastpage}
\end{document}